\documentstyle[amssymb,epsfig,preprint,aps]{revtex}
\tightenlines
\begin{document}
\title{{\bf Optimal packing of polydisperse hard-sphere fluids }}
\author{{\bf Junfang Zhang, Ronald Blaak,
Emmanuel Trizac\thanks{Present address:
Laboratoire de Physique Th\'eorique et Hautes Energies (URA D00063 du CNRS),
B\^atiment 211, Universit\'e  Paris-Sud,
91405 Orsay Cedex (France).},\\
Jos\'e A.\ Cuesta}$\thanks{
Permanent address: Grupo Interdisciplinar de Sistemas Complicados (GISC),
Departamento de Matem\'{a}ticas, Escuela Polit\'{e}cnica Superior,
Universidad Carlos III de Madrid, c/Butarque, 15, 28911-Legan\'{e}s, Madrid
(Spain).}${\bf \ and Daan Frenkel }}
\address{FOM Institute for Atomic and Molecular Physics, Kruislaan 407,\\
1098 SJ Amsterdam, The Netherlands}
\date{\today}
\maketitle

\begin{abstract}
We consider the effect of intermolecular interactions on the optimal
size-distribution of $N$ hard spheres that occupy a fixed total volume. When
we minimize the free-energy of this system, within the Percus-Yevick
approximation, we find that no solution exists beyond a quite low threshold (
$\eta \thickapprox 0.260$). Monte Carlo simulations reveal that beyond this
density, the size-distribution becomes bi-modal. Such distributions cannot
be reproduced within the Percus-Yevick approximation. We present a theoretical
argument that supports the occurrence of a non-monotonic size-distribution
and emphasizing the importance of finite size effects.
\end{abstract}

\section{Introduction}

Synthetic colloids are never perfectly monodisperse. Often, this
polydispersity is a drawback---for instance, polydispersity is a problem in
the preparation of high-quality colloidal crystals, that are
needed in photonic bandgap materials. However, occasionally, polydispersity
is desirable, because it allows us to achieve material properties that
cannot be realized with monodisperse colloids. For instance, monodisperse
colloidal systems can fill at most 74.05 \% of space in the crystalline
phase (regular close packing) and some 63\% in the liquid/glassy state
(random close packing). In contrast, colloids with a properly chosen
particle-size distribution can be made essentially space filling, both in
the crystalline solid (Appolonian packing) and in the liquid. In practice,
perfect space filling structures are never achieved because this requires an
infinite number of (predominantly small) particles per unit volume. Here, we
consider a somewhat simpler problem, namely the filling of a
given volume $V$ by a
fixed number of particles $N$, that occupy a prescribed
total volume fraction $\eta $.
We assume that the particles are free to exchange volume. As we have fixed
both the number and the total volume of the particles, the average volume
per particle is fixed---it defines the natural length-scale in the model.
Clearly, the Helmholtz free energy of the system will depend on the
non-fixed particle-size distribution. The distribution, however, is 
restricted by the two constraints of fixed number of particles and 
fixed total volume. We define the optimal size distribution to be the 
one that minimizes the Helmholtz free energy  under both
constraints. In the Monte Carlo simulations (performed in the
isothermal-isobaric ensemble) that we report
in this paper, we study the density dependence of the particle-size
distribution. We compare the simulation results for the size distribution
with an analytical estimate that is obtained by solving the Percus-Yevick
(PY) equation for an $N$-component hard-sphere mixture\cite{Salacuse}. Not
surprisingly, the PY equation works very well at low densities. However, the
theory breaks down at a surprisingly low density ($\eta \thickapprox $0.26).
Of course, the fact that an approximate theory fails at a given density,
does not imply that there is anything special going on in the system at that
density. Yet, our simulations indicate that there is---the size distribution
that was initially uni-modal, becomes bi-modal. We present
a theoretical argument supporting this scenario.

The remainder of this paper is organized as follows: in section \ref{simu}
we describe the constant-pressure Monte Carlo simulations. The Percus-Yevick
expression for the free energy of a system of polydisperse hard spheres with
variable size distribution is discussed in section \ref{idealentr} and
\ref{theory}.
Section \ref{sec:analytical} is devoted to the derivation of analytical
results turning useful when interpreting the simulation data.
The mechanism behind the
transition from uni-modal to bi-modal size distribution is also discussed.

\section{Monte Carlo Simulations}
\label{simu}

\begin{center}
\begin{figure}[h]
\vspace{4cm}
\epsfig{figure=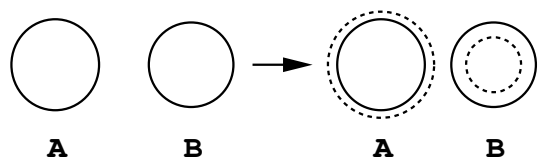,width=7cm,angle=0} 
\vspace{4cm}
\caption[a]{\label{poly_move} Zhang, Journal of Chemical Physics}
\end{figure}
\end{center}

Monte Carlo simulations were performed in the isothermal-isobaric
(constant $NPT$) ensemble \cite{FrenkelSmit}. This means that the
number of particles $N$, the pressure $P$ and the temperature $T$ are
fixed. We attempt three distinct types of trial moves. We change the
positions of the particles and allow the volume of simulation box to
fluctuate, in order to equilibrate with respect to the applied
pressure. Since we do not expect any crystalline order at low
pressures, a cubic box shape is maintained. The third type of move
is the one related to sampling the polydispersity of the
system. To this end, we select two particles at random, between which we
exchange an amount of volume drawn uniformly from the interval
$[-\Delta V_{max},\Delta V_{max}]$  (Fig. \ref{poly_move}). The
maximum volume change ${\Delta V_{max}}$ was chosen such that the
acceptance of a volume exchange move is between 35 and 50 \%.
The relative frequency of the three moves is given by $N:1:N/2$.
The initial configurations are made by $N$ monodisperse spheres on a
simple cubic lattice.

\begin{center}
\begin{figure}[h]
\epsfig{figure=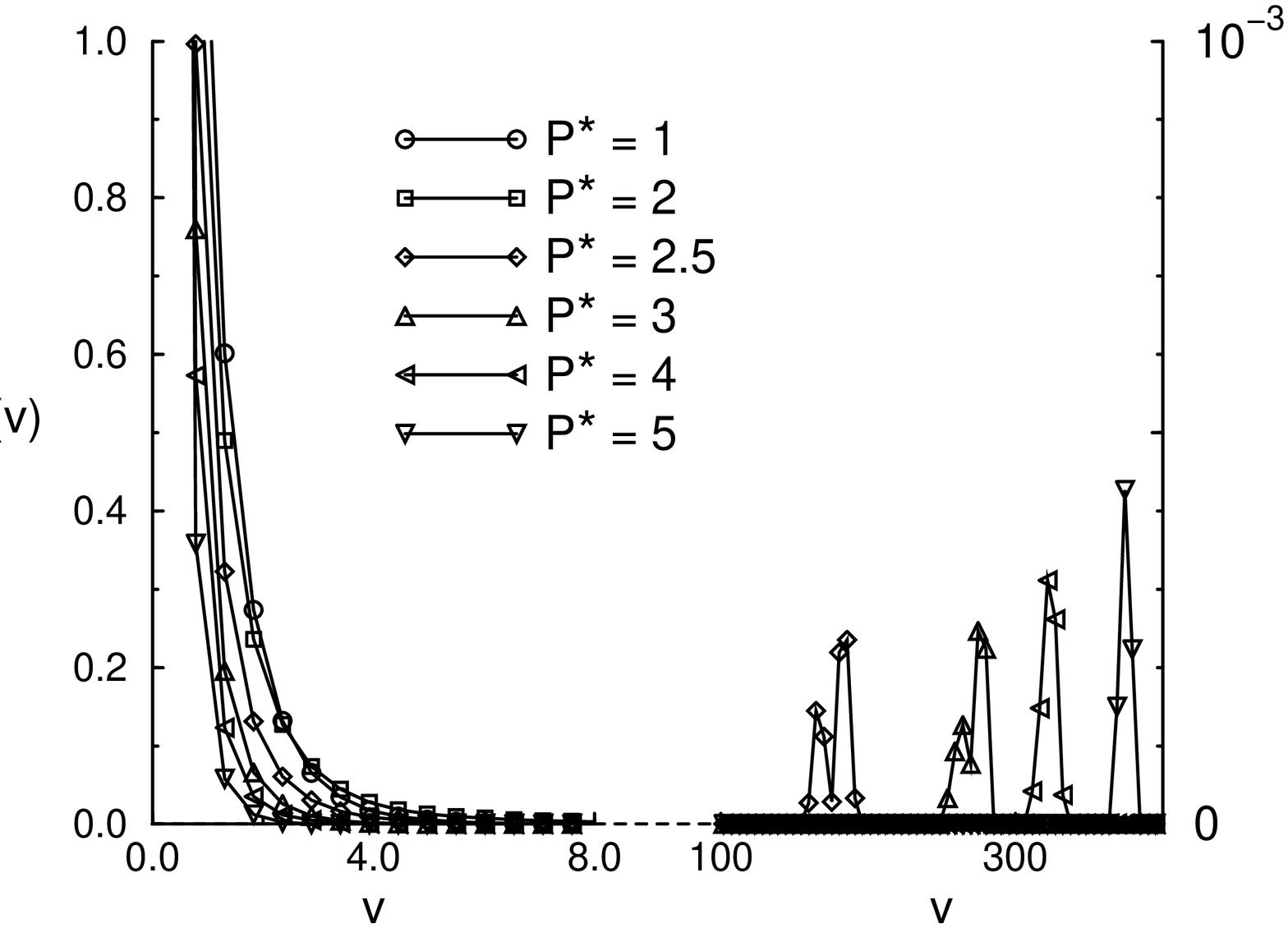,width=8cm,angle=0}
\vspace{4cm}
\caption[a]{\label{fig:uni-bi} Zhang, Journal of Chemical Physics}
\end{figure}
\end{center}

\begin{center}
\begin{figure}[h]
\epsfig{figure=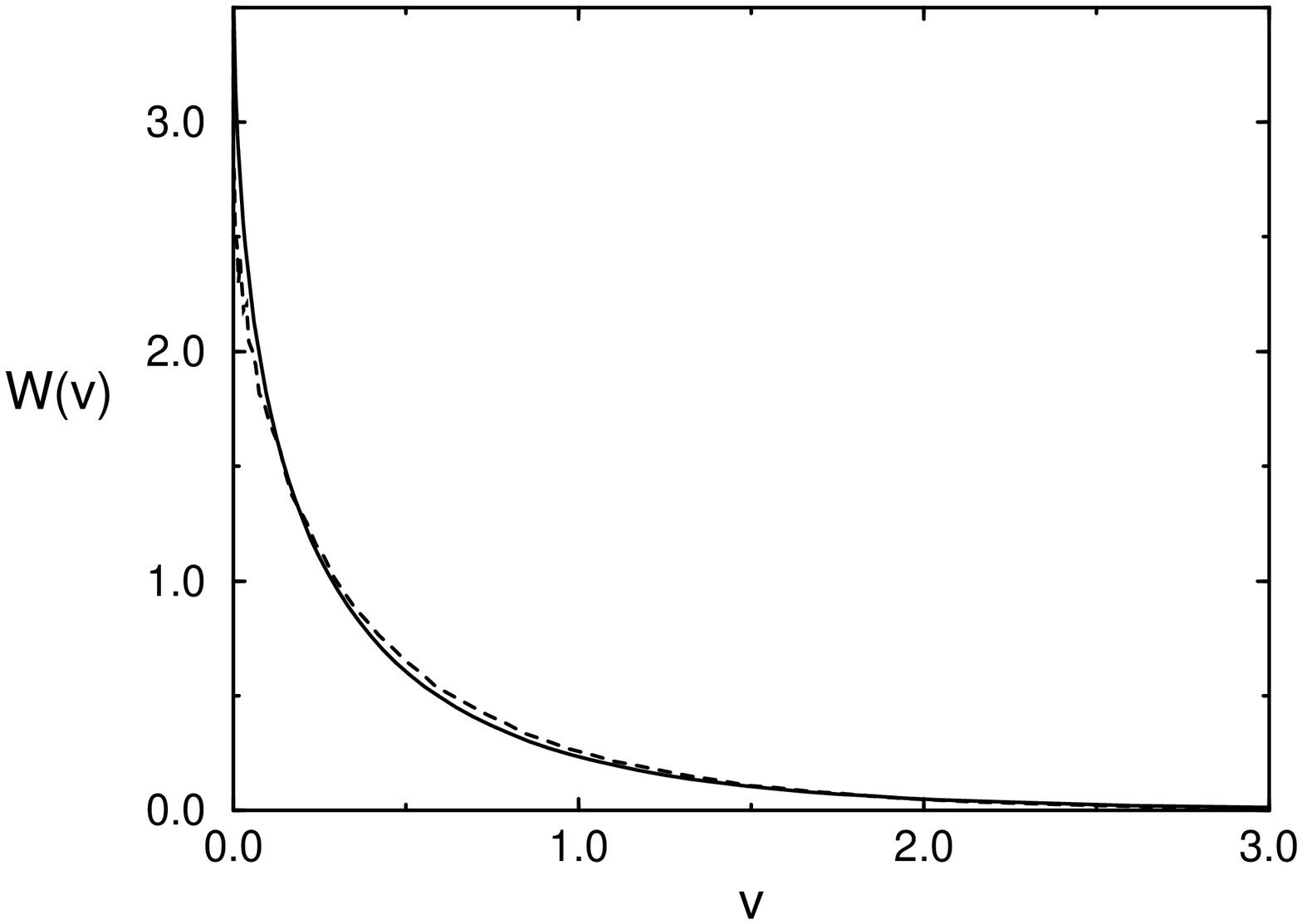,width=8.5cm,angle=0}
\vspace{4cm}
\caption{\label{prob_dist_pres_0.50} Zhang, Journal of Chemical Physics}
\end{figure}
\end{center}

\begin{center}
\begin{figure}[h]
\epsfig{figure=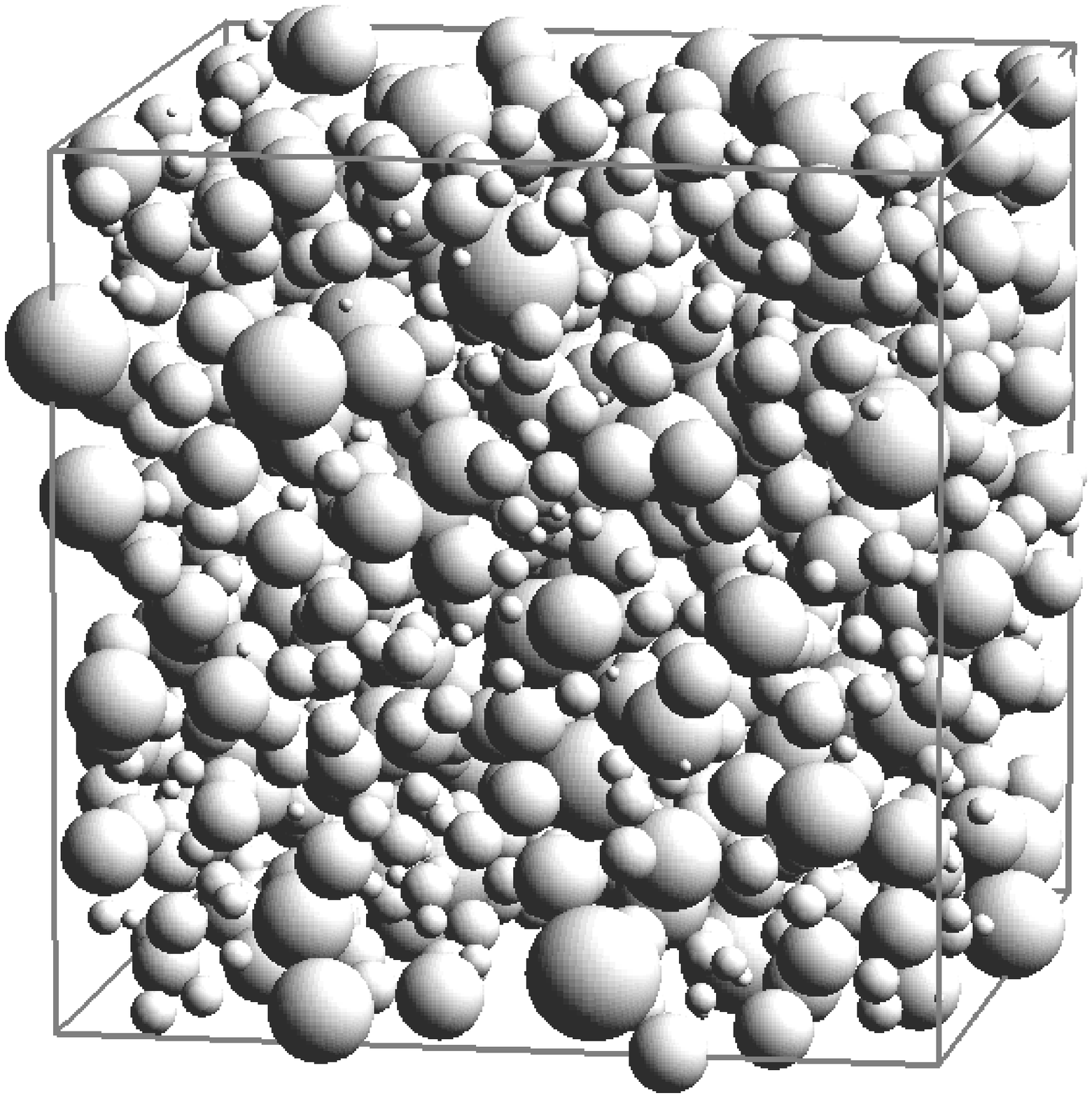,width=8.5cm,angle=0}
\vspace{4cm}
\caption[a]{\label{snap-low} Zhang, Journal of Chemical Physics}
\end{figure}
\end{center}

\begin{center}
\begin{figure}[h]
\epsfig{figure=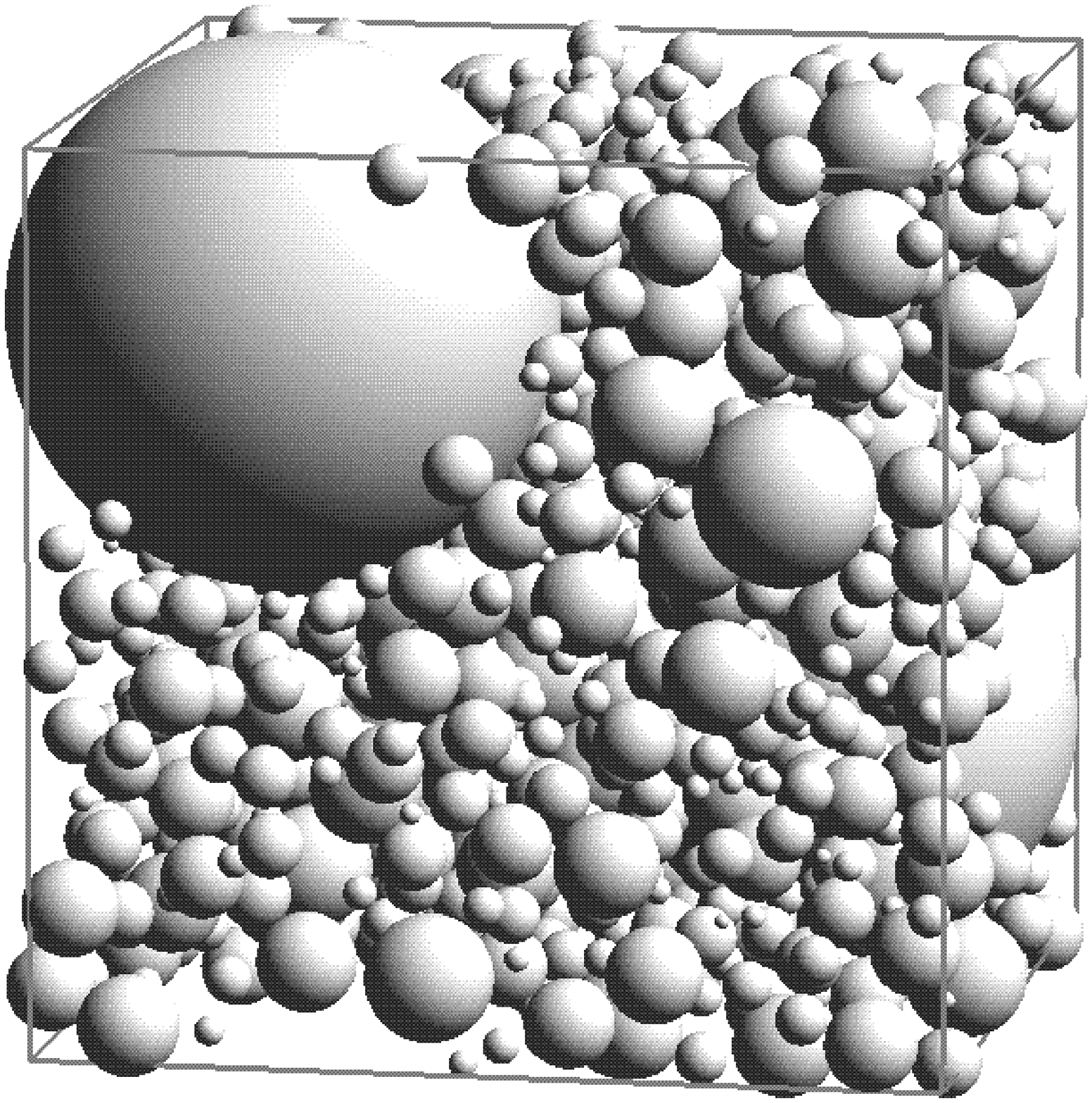,width=8.5cm,angle=0}
\vspace{4cm}
\caption[a]{\label{snap-high} Zhang, Journal of Chemical Physics}
\end{figure}
\end{center}

Simulations were performed for system sizes of $N$=512 or 1000, and
several different reduced pressures
${P}^{*}=(k_{B}T)^{-1}P\langle\sigma^{3}\rangle $, where $\sigma$ is
the diameter of particles, and $\langle\cdot \rangle $ denotes an
average over particle-size distribution. As $\langle \sigma^{3}\rangle
$ is fixed, we choose $\langle \sigma^{3}\rangle ^{1/3}$
to define the unit of length that we will use in the remainder of this
paper. We use $k_{B}T$ as our unit of energy. All other units that we
need, follow from these definitions. The equation of state [${P}^{*}$
as a function of ${\rho }^{*}(\equiv \rho \langle \sigma^{3}\rangle )$]
and the
particle size distribution function were determined in the simulations.
The results for the equation of state are shown in
Fig. \ref{eq_of_state}.
The simulation data have been collected in Table \ref{table-fequ1}.
At low
pressures the particle-size distribution function is a single-peaked
function with its maximum at $v=0$ (Figs. \ref{fig:uni-bi},
\ref{prob_dist_pres_0.50} and \ref{snap-low}). At higher pressures
(typically, $P^*>2.0$), the particle-size distribution develops a second
peak.
Actually, this second peak is quite small ({\it i.e.}
only a small fraction of
all
particles becomes ``large''). However, these particles contribute
appreciably to the total volume fraction (Fig. \ref{snap-high}).
Depending on the pressure this contribution
can get as large as 75\%.

The formation of big particles in these MC simulations is a rather
slow ``dynamical'' process. In order to speed up calculations, we did
additional simulations, in which we started with a bi-disperse
distribution, with one or several big particles containing 99\% of the
total volume occupied by the spheres, surrounded by a sea of small
particles containing the remaining volume.
In the 512 particle system, only one or two big particles remain for
the lower pressures ($P^*=2.5$ and $P^*=3.0$). For higher pressures
the number of big particles can stabilize at higher values as well.
For the 1000 particle system the maximum number of big particles
observed at the lower pressures is three. In addition, the size of
these big particles is not the same. It is not clear whether this
suggests a further possible fractionation or that it is a consequence
of the slow equilibration and that one or more of the big
particles are still shrinking.

Below, we discuss these simulation results in the context of the
relevant theoretical predictions, but first let us stop to make
certain considerations on the ideal entropy associated with this system.

\section{Ideal entropy of a polydisperse system}
\label{idealentr}

Strictly speaking, the ideal entropy of a polydisperse
system is infinite \cite{Salacuse}. In a multicomponent
system such an entropy is exactly given by
\begin{equation}
-Nk_B\sum_iw_i\ln(\Lambda_i^3\rho w_i)  \, ,
\label{ideal1}
\end{equation}
where $w_i$ is the molar fraction of species $i$ and
$\Lambda_i$ its thermal wavelength. The usual path
towards the entropy of a polydisperse system \cite{Salacuse}
(or for that matter, towards the entropy of continuous
signals in Information Theory \cite{reza}) is to
classify the species into ``boxes'' according to a
certain property which distinguishes them (diameter,
volume, molecular weight, etc). If $x$ denotes such a
property species $i$ will denote the box having particles
with $x$ between $i\Delta x$ and $(i+1)\Delta x$, for
a given $\Delta x$ which defines the boxes. If $W(x)$
denotes the probability density of a particle having
the value $x$ for that property, then $w_i=W(x_i)\Delta x$
where $x_i$ is a typical value of the $i$th box. Then Eq.
(\ref{ideal1}) adopts the form
\begin{equation}
-Nk_B\sum_iW(x_i)\Delta x\ln(\Lambda_i^3\rho W(x_i)/\Delta x)
\, .
\label{ideal2}
\end{equation}
The entropy of the polydisperse system is obtained from Eq.
(\ref{ideal2}) taking the limit $\Delta\to 0$, and we can
clearly see that, besides obtaining the usual expression
\cite{Salacuse,reza}
\begin{equation}
S_{\rm poly}=-NK_B\int dx\,W(x)\ln[\Lambda(x)^3W(x)] \, ,
\label{idealpoly}
\end{equation}
there is also a divergent $-\ln\Delta x$, which is simply taken
as a ``constant'' ignorable term.

But expression (\ref{idealpoly}) is not well defined. Suppose
we simply change coordinates to label the species from $x$ to
$y$ (say, from the diameter to the volume). Then the probability
density in the new variable will be $\widetilde W(y)=W(x)\left|
\frac{dx}{dy}\right|$. It is straightforward to show that 
in the new labeling the entropy becomes
\begin{equation}
S_{\rm poly}=-NK_B\int dy\,\widetilde W(y)
\ln[\Lambda(x(y))^3\widetilde W(y)]+Nk_B \int dy\,
\widetilde W(y)\ln\left|\frac{dy}{dx}\right|     \, ,
\label{newideal}
\end{equation}
which is different of what we would have obtained had we began
with the labeling $y$. 

This is a well known fact in Information Theory \cite{reza}.
In the study of fluid equilibria of {\em quenched} polydisperse
systems this fact turns out to be irrelevant because the new
term simply adds the same constant to both sides of the equilibrium
equations \cite{kincaid}. However when studying {\em annealed}
polydispersity this results tells us that the labeling is crucial
and has to be dictated by the physical process underlying the
polydispersity. In our case the Monte Carlo movements described
in Section \ref{simu} are a large scale description of a hypothetical 
microscopic system of tiny particles of exactly the same size
distributed among $N$ aggregates of a variable number of particles.
The constant volume constraint would correspond to the conservation
of the number of tiny particles, and the natural labeling of
the aggregates would be the number of tiny particles which form
it. As this number is proportional to the volume of the aggregate,
in the continuum description is the volume, instead of the diameter,
what turns out to be the natural labeling.

Notice that we could have described another model in which the tiny
particles aggregated into spherical surfaces. In that case it would
be the total surface what would be preserved and the natural labeling
of aggregates would be their respective surfaces. As we will discuss
in the conclusions, the physics of this model would be completely 
different.

\section{Percus-Yevick Theory}
\label{theory}

The Percus-Yevick equation for an $n$-component hard-sphere
mixture can be solved analytically, to yield the following equation of state
\cite{Salacuse}:
\begin{equation}
\frac{\pi }{6}P^*=\frac{\xi _{0}}{1-\xi _{3}}+\frac{3\xi _{1}\xi _{2}}{(1-\xi
_{3})^{2}}+\frac{3\xi _{2}^{3}}{(1-\xi _{3})^{3}}.  \label{PY-eos}
\end{equation}
The $j$-th moment of the particle-size distribution $\xi _{j}$ is defined as
\begin{equation}
\xi _{j}=\frac{\pi}{6}\sum_{i}\rho _{i}\left(\frac{6v_{i}}{\pi}
\right)^{j/3},  \label{xi-j}
\end{equation}
where $\rho _{i}=N_{i}/V$, the index $i$ is used to denote the different
particle species, and $v_{i}$ is the volume of the $i$-th species.

Equation (\ref{PY-eos}) is also valid for a continuous size distribution, in
which case the sum in Eq. (\ref{xi-j}) is replaced by an integral. The
corresponding expression for the chemical potential of a species with radius
$R$ is\cite{Baxter}
\begin{equation}
\mu^* =\ln\left[\rho \Lambda ^{3}W(v)\right]-
\ln (1-\xi _{3})+\frac{6\xi _{2}R
}{(1-\xi _{3})}+\frac{12\xi _{1}R^{2}}{(1-\xi _{3})}+\frac{18\xi
_{2}^{2}R^{2}}{(1-\xi _{3})^{2}}+\frac{4\pi }{3}P^* R^{3}  \label{PY-mu}
\end{equation}
where $\Lambda $ is the de-Broglie thermal wavelength, $\sqrt{h^{2}/(2\pi
mkT)}$, and $W(v)$ is the probability density to find a particle with a
volume around $v=(4\pi/3)R^3$. The pressure $P^*$ is given by
Eq. (\ref{PY-eos}).

In an (NPT) description, the Gibbs free energy of the system
fulfilling the constraints, must be at a minimum. The conservation of
the number of particles and of the solid volume fraction, imply that
$W(v)$ must be of the form:
\begin{equation}
W(v)=\exp \left\{ \sum_{i=0}^{3}\alpha _{i}R^{i}\right\} ,  \label{g(nu)-2}
\end{equation}
where
\begin{eqnarray}
\alpha _{1} &=&-\frac{6\xi _{2}}{1-\xi _{3}},  \label{alpha-1} \\
\alpha _{2} &=&-12(\frac{\xi _{1}}{1-\xi _{3}}+\frac{3\xi _{2}^{2}}{2(1-\xi
_{3})^{2}}).  \label{alpha-2}
\end{eqnarray}
The coefficients $\alpha _{0}$ and $\alpha _{3}$ are determined by the
constraints that the number of particles and the solid volume fraction are
fixed. Note that all ~$\xi _{i}$ $(i=1,2,3)$ are positive. Moreover, $\xi
_{3}$ is equal to the volume fraction $\eta $, and is therefore necessarily
less than one. Hence, $\alpha _{1}$ and $\alpha _{2}$ are always negative.
The last coefficient, $\alpha _{3}$, should be negative or zero, because
otherwise the particle-size distribution cannot be normalized. Since $\alpha
_{1}$, $\alpha _{2}$ and $\alpha _{3}$ are always negative the Percus-Yevick
equation predicts that $W(v)$ is a monotonically decreasing function of $v$.
This implies that the size-distribution given by Eq. (\ref{g(nu)-2}) can never
be bi-modal.

Note that these conclusions also hold for the more accurate
equation of state of Mansoori {\it et al.}~\cite{Mansoori}.
This equation adds an
extra term to the pressure given in
Eq. (\ref{PY-eos}) depending on $\xi _{3}$.
Thus Eq. (\ref{PY-mu}) is the same with the new expression for $P$---which
turns out to be irrelevant because the $R^{3}$ term is controlled by the
Lagrange multiplier associated with the constraint on the total solid
particle volume. In other respect, the analysis for the Mansoori
equation-of-state is identical to that for PY.

In practice, we solve Eq. (\ref{g(nu)-2}) numerically. To this end, we
represent $W(v)$ as a histogram. Initially, the value of $W(v)$ in the
different bins is assigned an arbitrary non-negative value, compatible with
the constraint that $W(v)$ is normalized and that $\langle v\rangle$ is
fixed. We fix the density at the desired value. We determine the optimal $
W(v)$ using the following scheme: we select a bin (say $i$) at random and
change the value of $W(v_{i})$ by a random amount $\Delta W$, distributed
uniformly in the interval $[-\Delta W_{max},\Delta W_{max}]$. We first check
if the new value $W(v_{i})$ is non-negative. If it is, we satisfy the
constraints by scaling the width of all bins and the height of the function
by two appropriately chosen factors. We then compute all moments $\xi _{j}$,
the pressure and the free energy, and we check if the Helmholtz free energy
is smaller than the previous one. If it is, we accept the new value for $
W(v_{i})$, otherwise we reject it. We repeat the procedure until the free
energy no longer decreases. We have verified that $W(v)$ is indeed of the
form given by Eqs. (\ref{g(nu)-2}) through (\ref{alpha-2}). Figure
\ref{prob_dist_pres_0.50} shows a comparison of the
PY estimate for $W(v)$, determined in this way, with the results of the full
Monte Carlo simulations. We find that $\alpha _{3}$ is a monotonically
increasing function of density. A comparison between simulation and PY
theory for the equation of state is shown in Fig. \ref{eq_of_state}. Note
that, in this figure, the PY solution terminates at pressure $
P_{c}\thickapprox 1.34$ (the cross in Fig. \ref{eq_of_state}). This is the
point where $\alpha _{3}$ becomes zero. Beyond this point we can no longer
find a solution for $W(v)$ that is of the form given by Eq. (\ref{g(nu)-2}).
In Appendix \ref{app:py}, we consider the breakdown of the PY theory in
more detail and  obtain the
packing fraction beyond which the PY approximation breaks down:
$\eta_{c}=0.260$, the corresponding pressure being
$P^*_{c}=1.343$.
This breakdown of the PY equation at a relatively low density is
surprising, as the PY equation works well up to quite high densities for
{\em fixed} particle size distributions \cite
{JacksonRowlinson,VosFrenkel,KranendonkFrenkel}. That it breaks down
regardless the accuracy of the equation of state can be inferred from the
fact that Mansoori {\it et al.}'s equation of state undergoes exactly the same
breakdown, though for slightly different values of $\eta $ and $P$. Besides,
from the analysis that we have carried out, it can be seen that a similar
breakdown will appear for any other theory yielding an equation of state
depending only on $\xi _{i}$, $i=0,1,2,3$.

\section{Analytical results}
\label{sec:analytical}
In this section, we derive a theoretical bound for the pressure
of the polydisperse system, providing the equation of state
at the packing fraction where the size distribution becomes bi-modal.
To this end, we work in the ($NVT$) ensemble and take advantage of the
extremality of the Helmholtz free energy under the constraints
of constant $N$ and $\eta$: the ``grand potential''
\begin{equation}
{\cal R} = {\cal F}\{W\} - {\cal L}_0 \int W(v) \, dv - {\cal L}_1  \int
v W(v) \, dv,
\end{equation}
where ${\cal L}_0$ and ${\cal L}_1$ are Lagrange multipliers,
has to be minimum for the optimal size distribution. In the above
relation, the free energy functional ${\cal F}$ can be cast into
the usual ideal and excess contributions
\begin{equation}
{\cal F}\{W\} = N k_B T\, \int dv \, W(v) \left[ \ln \left(\Lambda^3
\rho W(v)
\right) -1 \right] + {\cal F}_{\hbox{\scriptsize excess}}\{W\}.
\end{equation}

We attempt the following change in the system: the volume of a given particle
$v_0$ is increased by an amount $\delta v_0$, before a rescaling of all
volumes by a factor $\lambda$ ({\it e.g.} $v\to \lambda v$) such that
the overall volume change vanishes. This imposes:
\begin{equation}
\lambda = 1-\frac{\delta v_0}{N \langle v \rangle} + {\cal O}\left(
[\delta v_0]^2\right).
\end{equation}
The effect of the expansion of particle $v_0$ on the size-distribution can
be written:
\begin{equation}
\delta W(v) = \frac{1}{N}\,\,\left[\, \delta \left( v-v_0-\delta v_0\right)
\,-\, \delta\left(v-v_0\right) \, \right],
\end{equation}
where $\delta(...)$ denotes the Dirac distribution. The scaling
procedure affects $W$ according to
\begin{eqnarray}
\delta W(v) &=& \frac{1}{\lambda}\, W\!\left(\frac{v}{\lambda}\right) - W(v)\\
&=& \frac{\delta v_0}{N\langle v\rangle}\, \, \frac{d [v W]}{dv}
\,+\, {\cal O}\left([\delta v_0]^2\right).
\end{eqnarray}
The corresponding variation of the ideal contribution to ${\cal F}$
reads
\begin{equation}
\delta {\cal F}_{\hbox{\scriptsize id}} = \int dv\, \,
\frac{\delta {\cal F}_{\hbox{\scriptsize id}}\{W\}}{\delta W(v)} \,\,
\delta W(v),
\end{equation}
with the functional derivative
\begin{equation}
\frac{\delta {\cal F}_{\hbox{\scriptsize id}}\{W\}}{\delta W(v)} =
N k_B T \, \ln[\Lambda ^3 W(v)].
\end{equation}
We then get the entropic term
\begin{eqnarray}
\delta {\cal F}_{\hbox{\scriptsize id}} &=& k_B T
\int dv \, \ln[\Lambda ^3 W] \left[
\delta \left( v-v_0-\delta v_0\right)-\delta\left(v-v_0\right) +
\frac{\delta v_0}{\langle v\rangle}\,
\frac{d [v W]}{dv} \right]
\!\!\!\!\!\!\!\!\!     \\
&=& k_B T \left\{ \frac{W'(v_0)}{W(v_0)} +\frac{1}{\langle v\rangle}\right\}
\delta v_0,
\end{eqnarray}
where $W'$ is the derivative of $W$.

\begin{center}
\begin{figure}[h]
\epsfig{figure=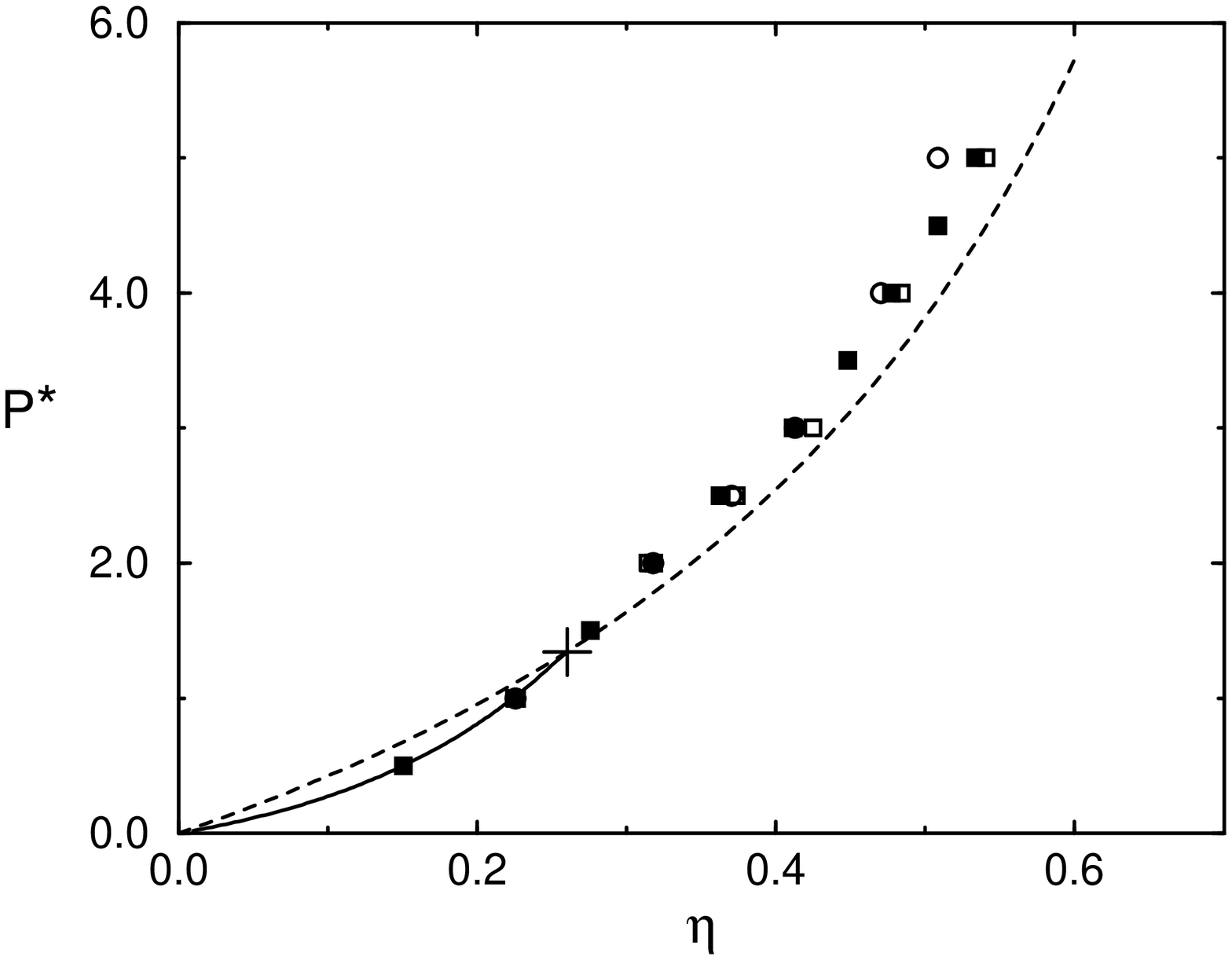,width=8.5cm,angle=0}
\vspace{4cm}
\caption{\label{eq_of_state} Zhang, Journal of Chemical Physics}
\end{figure}
\end{center}

The variation of the excess free energy reduces to the reversible work
needed to perform the transformation, and is derived in appendix
\ref{app:work}:
\begin{equation}
\frac{\delta W_{\hbox{\scriptsize rev}}}{\delta v_0} = \rho k_B T
\int dv\, W(v)\,\, g\left(\frac{\sigma_0 + \sigma}{2}\right) \left(1+
\frac{\sigma}{\sigma_0}\right)^2 \,-\,
\frac{P_{\hbox{\scriptsize excess}}}{\eta},
\label{eq:Wrevtout}
\end{equation}
where $g(\sigma_0/2+\sigma/2)$ denotes the radial distribution
function evaluated at contact between species of diameters
$\sigma_0$ (having volume $v_0$) and $\sigma$ (having volume $v$).
When $\sigma_0 \gg \langle \sigma \rangle$, we can replace
the density at the surface of particle $v_0$ by that
at a planar wall, and Eq. (\ref{eq:Wrevtout}) becomes
\begin{equation}
\delta W_{\hbox{\scriptsize rev}} = \left[ P -
\frac{P_{\hbox{\scriptsize excess}}}{\eta} \right] \delta v_0.
\end{equation}
In this limit $v_0 \gg \langle v \rangle$
\begin{equation}
\frac{\delta {\cal R}}{\delta v_0} =
\frac{\delta {\cal F}}{\delta v_0} = k_B T
 \left\{ \frac{W'(v_0)}{W(v_0)} +\frac{1}{\langle v\rangle}\right\}
+  P - \frac{P_{\hbox{\scriptsize excess}}}{\eta}.
\label{eq:rouge}
\end{equation}
For the optimal size distribution, $\delta F$ vanishes
so that
\begin{equation}
k_B T \, \frac{W'(v_0)}{W(v_0)} + \frac{1\!-\!\eta}{\eta} \left[
\frac{2\,\rho \, k_B T}{1-\eta} -P \, \right] = 0.
\label{eq:equilW}
\end{equation}
Assuming $W(v)$ to be a normalizable distribution, $W'(v)$
has to be negative for large arguments, which sets the upper bound:
\begin{equation}
P\, <\,2\, \frac{\rho k_B T}{1-\eta} \qquad \hbox{or} \qquad P^* \, < \,
\frac{12\, \eta}{\pi (1-\eta)}
\label{eq:pbound}
\end{equation}
for the rescaled pressure.
For low packing fractions ($\eta < \eta_c$) where the PY solution
is available, the above inequality is fulfilled (Fig.
\ref{eq_of_state}). At the threshold $\eta=\eta_c$ where the second
polydispersity peak appears, $W'$ changes sign which means
\begin{equation}
P^*_c = \frac{12\, \eta_c}{\pi(1-\eta_c)}.
\label{eq:upperbound}
\end{equation}
The above relation is remarkably well obeyed within the PY approximation
(see the data of section \ref{theory} or figure \ref{eq_of_state}:
the PY expression crosses the line given by Eq. (\ref{eq:pbound})
exactly at $\eta_c$). For $\eta > \eta_c$, the upper bound is violated
by the simulation results reported in Table \ref{table-fequ1} and figure
\ref{eq_of_state}. However, the data suggest non negligible
finite-size effects: increasing $N$ shifts the pressure
closer to the theoretical bound.
Besides, starting from bi-disperse initial conditions
(cf the procedure described in section \ref{simu}), supposed to be
closer to the expected polydispersity,
has the same effect. According to expressions (\ref{eq:rouge}) and
(\ref{eq:equilW}), the violation of Eq. (\ref{eq:pbound}) means
that $\delta {\cal R} = \delta {\cal F} <0$ for
$\delta v_0 >0$,
so that the biggest particle tends
to expand. Its growth is however necessarily limited
by the length $L$ of the simulation box.
This is supported by the observation that,
even for the largest
system investigated ($N=10^3$ particles), the size of the
biggest particle obtained is determined by $L$
($\sigma_{\hbox{\scriptsize biggest}} > L/3$,
irrespective of the packing fraction, see for example
figure \ref{snap-high}). This suggests that
system sizes that would presumably allow the system to
reach thermodynamic equilibrium (and fulfill inequality
(\ref{eq:pbound})) are numerically out of reach.
Consequently, the question of the extensivity/intensivity
of the number of large particles cannot be addressed
by simulations; a theoretical investigation seems to require
the detailed knowledge of the interfacial free energy between
``large'' and ``small'' species.
At this stage, we cannot tell whether a true phase transition is
associated with the occurrence of the second peak in the particle-size
distribution. The available data certainly do not rule out this
possibility.

Finally, the integration of Eq. (\ref{eq:equilW})
yields the tail of the optimal size-distribution:
\begin{equation}
\ln W(v) \propto -v/\langle v\rangle
\qquad \hbox{for} \qquad v \gg \langle v \rangle.
\end{equation}
Equation (\ref{g(nu)-2}) obeys this relation, which cannot be
tested against simulation results because of the lack
of statistics for very large particles (not more than 5 in a typical run).

\section{Conclusion}

At this
stage, we can only speculate what will happen at larger $N$
and/or larger densities.
Conceivably, once the volume-fraction of the large particles exceeds a
certain threshold, proliferation of still larger particles can occur, and so
on, until eventually an ''Appolonian'' packing of the liquid is achieved.
The theoretical analysis of this scenario is non-trivial, as the small
particles now induce attractive depletion forces between the large
particles. Unfortunately, the systems that we can conveniently study by
simulation are too small to allow us to investigate this regime.

We stress that the specific model we have chosen to study
is somewhat arbitrary. For instance, rather than fixing the number of
particles, one might have chosen to fix the total surface area of the
particles. The latter constraint would be logical if one aims to model the
size distribution of droplets covered with a fixed amount of surfactant. In
addition, we assume that the surface free energy of the spheres is
negligible. Again, this constraint can be removed. We hope that the rather
surprising results of the present study will stimulate research into these
related models.

\section*{Acknowledgments}

We thank Bela Mulder for critical discussions.
The work of the FOM Institute is part of
the research program of FOM and is made possible by financial support from
the Netherlands Organization for Scientific Research (NWO). Junfang Zhang
acknowledges financial support from Chinese government. Jos\'e Cuesta
participated in this work during a stay at AMOLF financed by the project
no.\ PR95-558 from the Direcci\'on General de Ense\~nanza Superior (DGES).
His work is also part of the DGES project no.\ PB96--0119. He also wants to
thank the hospitality found at AMOLF, specially from Daan Frenkel's and Bela
Mulder's groups.

\appendix
\section{}
\label{app:py}

Let us consider the range of densities where a solution of the Percus-Yevick
equation is possible. As stated in section \ref{theory},
it is essential that $\alpha _{3}$
be non-positive. Hence, the pressure at which $\alpha _{3}=0$ defines the
end point of the theory. To locate this point, consider the form of the
solution at the point where $\alpha _{3}=0$. Then the distribution function
reduces to
\[
W(v)=\exp \left\{ \sum_{i=0}^{2}\alpha _{i}R^{i}\right\} ,
\]
where $R$ hereafter denotes a reduced radius measured in units of
$\langle \sigma^3 \rangle^{1/3}$.
Using the two constraints for normalization and for the average volume of
the particles, we can express the coefficients $\alpha _{0}$ and $\alpha
_{2} $ in terms of $\alpha _{1}$. If we combine Eqs. (\ref{alpha-1}) and (\ref
{alpha-2}) to eliminate $\xi _{3}$, we obtain
\begin{eqnarray}
\alpha _{2} &=&2\alpha _{1}\frac{\xi _{1}}{\xi _{2}}-\frac{1}{2}\alpha
_{1}^{2}  \label{a1-a2} \\
&=&\alpha _{1}\frac{\langle R\rangle }{\langle R^{2}\rangle }-\frac{1}{2}
\alpha _{1}^{2},  \nonumber
\end{eqnarray}
where the ratio of the moments $\xi _{1}$/$\xi _{2}$ does not depend
explicitly on the density $\rho $ or on $\alpha _{0}$, but it only depends
on $\alpha _{1}$ itself. In the second line we have used the definition
\begin{equation}
\langle R^{i}\rangle =\int_{0}^{\infty }R^{i}W(v)4\pi R^{2}dR.
\end{equation}
But we have another relation between the moments of the particle-size
distribution: partial integration of $\int R^{n}W(v)dv$ yields
\begin{equation}
\left[ R^{n+1}W(v)\right] _{0}^{\infty }=\int_{0}^{\infty }\left( 2\alpha
_{2}R^{n+2}+\alpha _{1}R^{n+1}+(n+1)R^{n}\right) W(v)dR=0
\end{equation}
for $n\geq 0$. This leads to the identities
\begin{eqnarray}
\langle R^{3}\rangle &=&\frac{-\alpha _{1}}{2\alpha _{2}}\langle
R^{2}\rangle -\frac{4}{2\alpha _{2}}\langle R\rangle , \\
\langle R^{2}\rangle &=&\frac{-\alpha _{1}}{2\alpha _{2}}\langle R\rangle -
\frac{3}{2\alpha _{2}}\langle R^{0}\rangle .
\end{eqnarray}
This allows us to write $\langle R\rangle $ and $\langle R^{2}\rangle $ as a
function of $\alpha _{1}$, $\alpha _{2}$, $\langle R^{0}\rangle $ and $
\langle R^{3}\rangle $; i.e.
\begin{eqnarray}
\langle R\rangle &=&\frac{-3\alpha _{1}\langle R^{0}\rangle +4\alpha
_{2}^{2}\langle R^{3}\rangle }{\alpha _{1}^{2}-8\alpha _{2}}, \\
\langle R^{2}\rangle &=&\frac{12\langle R^{0}\rangle -2\alpha _{1}\alpha
_{2}\langle R^{3}\rangle }{\alpha _{1}^{2}-8\alpha _{2}}.
\end{eqnarray}
But $\langle R^{0}\rangle =1$ and $\langle R^{3}\rangle =1/8$; hence
\begin{equation}
\frac{\langle R\rangle }{\langle R^{2}\rangle }=\frac{-3\alpha {1}+\frac{1}{2
}\alpha _{2}^{2}}{12-\frac{1}{4}\alpha _{1}\alpha _{2}}.
\end{equation}
Substituting this expression in Eq. (\ref{a1-a2}), we can eliminate $\langle
R\rangle /\langle R^{2}\rangle $ to obtain
\begin{equation}
\alpha _{2}+\frac{1}{2}\alpha _{1}^{2}=\alpha _{1}\frac{-3\alpha
{1}+\frac{1}{2}\alpha _{2}^{2}}{12-\frac{1}{4}\alpha _{1}\alpha _{2}}.
\label{self-consist1}
\end{equation}
For what follows, it is convenient to introduce two new variables
$f$ and $c$
\begin{eqnarray}
\alpha _{1} &=&-fc,  \label{a1fca2ff} \\
\alpha _{2} &=&-f^{2};  \nonumber
\end{eqnarray}
then
\begin{equation}
\langle R^{i}\rangle =\frac{1}{f^{i+3}}\langle R^{i}\rangle _{f=1}.
\end{equation}
We can use this relation to express $f$ as function of $c$. We use the fact
that the average volume per particle is fixed, to rewrite
\begin{eqnarray}
\frac{\langle R^{3}\rangle }{\langle R^{0}\rangle } &=&\frac{1}{8}=\frac{1}{
f^{3}}\frac{\langle R^{3}\rangle _{f=1}(c)}{\langle R^{0}\rangle _{f=1}(c)}
\nonumber \\
&\equiv &\frac{1}{f^{3}}Y(c),
\end{eqnarray}
where the second line defines the function $Y(c).$ Hence
\begin{equation}
f=2Y^{1/3}(c)
\end{equation}
As $Y(c)$ can be expressed explicitly in terms of error functions, we now
know $f$ as a function of $c$. Equation (\ref{a1fca2ff}) allows us to
express both $\alpha _{1}$ and $\alpha _{2}$ as explicit functions of $c$.
We can then use Eq. (\ref{self-consist1}) to determine $c$ numerically. We
find that this equation has a unique solution. Once the value of $c$ has
been determined, we know $\alpha _{1},\alpha _{2}$ and $\alpha _{0}$ (from
the normalization condition). Eq. (\ref{alpha-1}) finally yields the
packing fraction beyond which the PY approximation breaks down: $\eta
_{c}=0.260198$. The corresponding pressure is $P^*_{c}=1.343442$.

\section{}
\label{app:work}

We first note that the reversible work done by an operator
rescaling both particle volumes ($v_i\to \lambda v_i,\, \forall i$)
and container volume ($V\to V'= \lambda V $), is
\begin{equation}
\delta W_1 = -P_{\hbox{\scriptsize ideal}}\, \delta V =
-\rho k_B T \, \delta V, \qquad \hbox {where} \quad
\delta V = (\lambda-1) \,  V.
\end{equation}
In the transformation, the total volume $V_p$ of the particles
changes according to
\begin{equation}
\frac{\delta V_p}{V_p} =\frac{\delta V}{V} \quad \Longrightarrow \quad
\delta V_p \,=\, \eta \, \delta V.
\end{equation}
Keeping the particle volumes fixed and going back to the original
container volume ($V'\to V'/\lambda$) requires the reversible work
\begin{equation}
\delta W_2 = - P\, (V-V') = P\, \delta V.
\end{equation}
It is then straightforward to obtain the work associated with
a rescaling of particle volumes at constant accessible volume $V$:
\begin{equation}
\delta W_{v \to \lambda v} \,=\, \delta W_1 + \delta W_2\, =\,
\frac{P_{\hbox{\scriptsize excess}}}{\eta} \, \delta V_p,
\label{eq:aBw1}
\end{equation}
valid for all polydispersities $W(v)$.

In the remainder, the shall derive the work needed to grow
a particle $v_0$ by an amount $\delta v_0$ ($\delta v_0 =
\pi \sigma_0^2 \delta \sigma_0/2$). We assume the normalization
$\int W dv = 1$ to hold.
Consider species having
volumes between $v$ and $v+\delta v$ (diameters between $\sigma$
and $\sigma+\delta \sigma$). They exert a pressure
$\rho k_B T \, W(v) dv\,  g(\sigma_0/2+\sigma/2) $ on particle $v_0$,
involving
the radial distribution function at contact between species
$\sigma_0$ and $\sigma$. For the above pair $(\sigma_0,\sigma)$,
the excluded volume sphere has diameter $\sigma_0+\sigma$,
and sweeps a volume
\begin{equation}
\delta V_{\hbox{\scriptsize sweep}} = \pi \, (\sigma_0+\sigma)^2 \,
\frac{\delta \sigma_0}{2} = \left(1 + \frac{\sigma}{\sigma_0}\right)^2 \,
\delta v_0
\end{equation}
during the growth of particle $v_0$. Summing over all species
$v$, the work performed by the operator takes the form
\begin{equation}
\delta W_{\hbox{\scriptsize growth $\!v_0$}}   = \rho k_B T \,
\delta v_0 \, \int
d v \, W(v)\, g\left(\frac{\sigma_0+\sigma}{2}\right)\left(
1+\frac{\sigma}{\sigma_0}\right)^2.
\label{eq:aBw2}
\end{equation}
For the size modification considered in section \ref{sec:analytical},
the global volume change of the particles vanishes, such
that $\delta V_p + \delta v_0 = 0$. Summing the contributions
arising from Eqs. (\ref{eq:aBw1}) and (\ref{eq:aBw2}),
the results of Eq. (\ref{eq:Wrevtout}) is recovered:
\begin{equation}
\frac{\delta W_{\hbox{\scriptsize rev}}}{\delta v_0} = \rho k_B T
\int dv\, W(v)\, g\left(\frac{\sigma_0 + \sigma}{2}\right) \left(1+
\frac{\sigma}{\sigma_0}\right)^2 \, -\, \,
\frac{P_{\hbox{\scriptsize excess}}}{\eta}.
\end{equation}

Note that a similar argument can be invoked to compute the reversible
work needed to rescale all particle diameters, at constant $V$. After
integration of Eq. (\ref{eq:aBw2}) over all $W(v_0)\, dv_0$,
we obtain:
\begin{equation}
\delta W_{v \to \lambda v} \,=\, \rho k_B T
\int dv\, dv'\, W(v)\, W(v')\, g\left(\frac{\sigma + \sigma'}{2}\right)
(\sigma+\sigma')^2 \, \sigma\,\, \frac{\delta V_p}{\langle \sigma \rangle^3}.
\end{equation}
Inserting this result into Eq. (\ref{eq:aBw1}) provides the
equation of state for a polydisperse fluid of hard spheres
\begin{equation}
\frac{P}{\rho k_B T} = 1 + \eta
\int dv\, dv'\, W(v)\, W(v')\, g\left(\frac{\sigma + \sigma'}{2}\right)
(\sigma+\sigma')^2 \, \frac{\sigma}{\langle \sigma \rangle^3}.
\end{equation}
For a monodisperse fluid, $W(v) = \delta(v-v_0)$ and we recover
the well known relation
\begin{equation}
\frac{P}{\rho k_B T} = 1 + 4 \,\eta \, g(\sigma_0) .
\end{equation}

\newpage
\begin{table}
\begin{tabular}{l|l}
$\eta $ & $P^*$ \\ \hline
0.0052(3) & 0.01 \\ \hline
0.244(1)  & 1.00 \\ \hline
0.314(3)  & 2.00 \\ \hline
0.374(2)  & 2.50 \\ \hline
0.425(4)  & 3.00 \\ \hline
0.484(2)  & 4.00 \\ \hline
0.541(3)  & 5.00
\end{tabular}
\vspace{4cm}
\caption{Equation of state of polydisperse hard spheres, obtained
from MC simulations with 1000 particles. The estimated error in the
last digit of the packing fraction $\eta $ is indicated in
parentheses. \label{table-fequ1}}
\end{table}

\begin{center}
{\large FIGURE CAPTIONS}
\end{center}

\begin{enumerate}

\item
Schematic drawing of the Monte Carlo trial used to sample the
polydispersity. Of two randomly chosen particles the volume of one
is increased, while the other is decreased in volume by the same
amount.

\item
The equation of state of the polydisperse system. The solid line is
the PY prediction, which can not be extended beyond the cross. The
circles correspond to a system initially prepared monodisperse, while
the squares are final values in which we initially started with one
big particle. The solid squares and the open symbols are from a 512
and 1000 particle system respectively. The dashed line is the upper
bound (\ref{eq:upperbound}) for the pressure.

\item
The Monte Carlo results for the distribution of particle volumes,
$W(v)$, as a function of $v$, for several reduced pressures for a 1000
particle system. For $P^*>2.0$ the distribution develops a second
peak (with statistical noise)
at much larger volumes (note the change in scale on the
right-hand side of the figure). Although there are only one or several
of these big particles, they can contribute over 30\% of the total
volume of all particles. The diameters can get larger than a third of
the length of the simulation box.

\item
Comparison of the numerical results (dashed line) for the
particle-size distribution $W(v)$, with the corresponding prediction
of the PY theory (solid line). These results were obtained at a
relatively low pressure ($P^*=0.5$), which corresponds to a volume
fraction $\eta =0.151$. Note that at this density,
simulation and theory are in quite
good agreement.

\item
Snapshot of a typical configuration at reduced pressure $P^*=1.0$ and
volume fraction $\eta=0.244$.

\item
Snapshot of a typical configuration at reduced pressure $P^*=3.0$ and
volume fraction $\eta=0.425$. In this case there are two big
particles.

\end{enumerate}

\newpage

\newpage

\newpage

\newpage

\newpage


\begin{references}
\bibitem{Salacuse}  J. J. Salacuse and G. Stell, J. Chem. Phys. {\bf
77}, 3714 (1982).

\bibitem{Baxter}  R. J. Baxter, J. Chem. Phys. {\bf 52}, 4559 (1970).

\bibitem{reza} F. M. Reza, {\em An introduction to Information
Theory} (Dover, New York, 1994).

\bibitem{kincaid} J. A. Gualteri, J. M. Kincaid, and G. Morrison,
J. Chem. Phys. {\bf 77}, 521 (1982).

\bibitem{Mansoori}  G. A. Mansoori, N. F. Carnahan, K. E. Starling,
and T. W. Leland, J. Chem. Phys. {\bf 54}, 1523 (1971).

\bibitem{FrenkelSmit}  D. Frenkel and B. Smit, {\em Understanding
Molecular Simulation:From Algorithms to Applications} (Academic Press,
San Diego, 1996). 

\bibitem{JacksonRowlinson}  G. Jackson, J. S. Rowlinson, and F. van
Swol, J. Phys. Chem. {\bf 91}, 4907 (1987).

\bibitem{VosFrenkel}  D. Frenkel, R. J. Vos, C. G. de Kruif, and A. Vrij, J.
Chem. Phys. {\bf 84}, 4625 (1986).

\bibitem{KranendonkFrenkel}  W. G. T. Kranendonk and D. Frenkel, Mol. Phys.
{\bf 72}, 715 (1991).
\end{references}
\end{document}